\documentclass[twocol]{ametsoc}

\usepackage{text_shortcuts}
\usepackage{math_shortcuts}
\usepackage{phys_shortcuts}

\usepackage{textcomp}  

\journal{jas}
\bibpunct{(}{)}{;}{a}{}{,}

\title{Solsticial Hadley Cell ascending edge theory from supercriticality}

\authors{Spencer A. Hill\correspondingauthor{Spencer Hill, Lamont-Doherty Earth Observatory, 61 Route 9W, Palisades, NY 10964}}
\affiliation{Lamont-Doherty Earth Observatory, Columbia University, New York, New York}
\email{shill@ldeo.columbia.edu}

\extraauthor{Simona Bordoni}
\extraaffil{Department of Civil, Environmental and Mechanical Engineering (DICAM), University of Trento, Trento, Italy, and Division of Geological and Planetary Sciences, California Institute of Technology, Pasadena, California}

\extraauthor{Jonathan L. Mitchell}
\extraaffil{Department of Atmospheric and Oceanic Sciences, and Department of Earth, Planetary, and Space Sciences, University of California, Los Angeles}

\abstract{%
How far the Hadley circulation's ascending branch extends into the summer hemisphere is a fundamental but incompletely understood characteristic of Earth's climate.  Here, we present a predictive, analytical theory for this ascending edge latitude based on the extent of supercritical forcing.  Supercriticality sets the minimum extent of a large-scale circulation based on the angular momentum and absolute vorticity distributions of the hypothetical state were the circulation absent.  We explicitly simulate this latitude-by-latitude radiative-convective equilibrium (RCE) state.  Its depth-averaged temperature profile is suitably captured by a simple analytical approximation that increases linearly with \(\sin\lat\), where \(\lat\) is latitude, from the winter to the summer pole.  This, in turn, yields a one-third power-law scaling of the supercritical forcing extent with the thermal Rossby number.  In moist and dry idealized GCM simulations under solsticial forcing performed with a wide range of planetary rotation rates, the ascending edge latitudes largely behave according to this scaling.
}

\begin{document}
\maketitle

\section{Introduction}
\label{sec:intro}
Why does the shared, ascending edge of Earth's Hadley cells sit around 15\degr{} latitude in the summer hemisphere, instead of say 1.5\degr{} or at the summer pole?  Results from idealized general circulation model (GCM) simulations suggest that neither limit is as outlandish as may initially seem.  For the former, an \(O(1)\) increase in the surface-atmosphere system's thermal inertia timescale leaves the ascending branch insufficient time to migrate more than a few degrees off the equator before the insolation maximum moves back toward the opposite hemisphere \citep[\eg/][]{donohoe_effect_2014}.  For the latter, the insolation distribution that ultimately drives the general circulation maximizes at the summer pole, and an \(O(1)\) decrease in the planetary rotation rate yields nearly pole-to-pole solsticial Hadley circulations \citep[\eg/][]{williams_range_1982}.

Although increasing the system's thermal inertia (or hastening the annual cycle) pulls the solsticial ascending branch equatorward, decreasing it (or slowing the annual cycle) does not push the branch much poleward --- even in the limit of time-invariant solsticial forcing \citep[\eg/][]{faulk_effects_2017,zhou_hierarchy_2018,singh_limits_2019}.  This suggests the presence of a dynamical constraint emanating from the time-mean forcing at solstice.


Several theories exist of direct or indirect relevance to this fundamental property of the general circulation, but each is limited in one or more substantive ways.  The energetic framework for the position of the Intertropical Convergence Zone \citep[ITCZ; \eg/][]{kang_response_2008,schneider_migrations_2014} is diagnostic and not always accurate, even qualitatively \citep[\eg/][]{hill_theories_2019}.  The solsticial equal-area model \citep{lindzen_hadley_1988} is predictive but inaccurate over much of the relevant parameter space, even restricting to axisymmetric atmospheres for which it is strictly applicable \citep{hill_axisymmetric_2019}.\footnote{By diagnostic, we mean that the theory requires knowledge of one or more fields from the dynamically equilibrated state that is nominally being predicted.  By predictive, we mean that the theory requires knowledge only of fields related to the forcing, thereby yielding a true prediction of the dynamically equilibrated state.  Naturally, all else equal, a predictive theory is preferable.}
A recent theory for the ascending edge based on slantwise convective neutrality \citep{singh_limits_2019} is quantitatively accurate across the idealized GCM simulations against which it has been tested, but it is diagnostic.  Here, we will pursue an alternative, predictive theory based on the extent of supercritical radiative forcing.

A supercritical latitude is one at which, supposing no large-scale overturning circulation existed, the resulting state of latitude-by-latitude radiative-convective equilibrium (RCE) would possess impossible distributions of angular momentum and absolute vorticity \citep{plumb_response_1992,emanuel_thermally_1995} --- that is, distributions that violate Hide's theorem \citep{hide_dynamics_1969}.\footnote{Particularly in extratropical contexts, the term ``supercriticality'' is sometimes used in reference to the isentropic slope.  In this manuscript, however, supercriticality always refers to Hide's theorem.}  A large-scale overturning circulation must therefore span at minimum all supercritical latitudes \citep{held_nonlinear_1980}.  Recent studies using idealized dry, axisymmetric \citep{hill_axisymmetric_2019} and moist, eddying \citep{faulk_effects_2017,singh_limits_2019} GCMs explore the qualitative utility of the supercritical forcing extent as a predictor of the solsticial Hadley cell extent as planetary rotation rate is varied.  But they fell short of deriving a closed, analytical expression for the solsticial supercritical forcing extent.

An attractive feature of the supercritical forcing extent is that its interpretation as setting the minimum extent of a large-scale circulation holds equally for axisymmetric and zonally varying atmospheres: by definition, RCE implies the absence of any large-scale circulation, and therefore over those latitudes where RCE cannot be sustained \emph{some} circulation has to emerge.  At the same time, it does not specify the nature of the large-scale circulation that emerges, in particular whether even Hadley like at all or instead strongly macroturbulent as in the extratropics.  Using the supercritical forcing extent as a theory specifically for the Hadley cell ascending edge, therefore, entails some additional empirical justification.  A more beneficial corollary of this dynamical agnosticism, though, is that the supercritical forcing extent's validity does not depend on the resulting Hadley cells being in one of the two limiting regimes of the zonal momentum budget --- angular momentum conserving or eddy dominated.  Such limit-based approaches will always be incomplete for the simple reason that Earth's solsticial Hadley cells do not consistently adhere to one or the other limit \citep[\eg/][]{schneider_general_2006,bordoni_monsoons_2008}.

For annual-mean forcing, an analytical expression for the extent of supercritical forcing has been known for decades thanks to \citet{held_nonlinear_1980}, who assume an RCE depth-averaged temperature profile varying simply as \(\sin^2\lat\), where \(\lat\) is latitude.\footnote{Of course, in the annual-mean the ascending edge will reliably sit near the equator (potentially as a double ITCZ straddling the equator), and the utility of the supercritical forcing extent is as a lower bound for the location of the poleward, descending Hadley cell edges.}   For solsticial forcing, then, a natural starting place is an analytical RCE profile that moves the global maximum of the RCE temperature field off the equator but retains the simple \(\sin^2\lat\) meridional dependence as in \citet{held_nonlinear_1980} --- namely, that presented by \citet[][hereafter LH88]{lindzen_hadley_1988}  In fact, a cruder \(\sin\lat\) approximation will prove adequate.

This paper addresses these issues by showing that:
\begin{itemize}
\item conceptually, supercritical forcing extent can constitute a meaningful theory for the solsticial Hadley circulation ascending latitude in zonally varying atmospheres, provided certain empirical claims are established (Section~\ref{sec:why-relevant});
\item the LH88 forcing usefully approximates latitude-by-latitude RCE under solsticial forcing with respect to fields relevant to the Hadley cells (Section~\ref{sec:rce-sims});
\item a simple, approximate analytical solution exists for the supercritical forcing extent at solstice based on the LH88 forcing (Section~\ref{sec:theory-summer}); and
\item the cross-equatorial Hadley cell extent obeys this simple scaling in previously reported moist idealized GCM simulations as well as newly performed dry idealized GCM simulations (Section~\ref{sec:sims}).
\end{itemize}
We then discuss how our theory relates to the aforementioned slantwise convective neutrality diagnostic (Section~\ref{sec:disc}) before concluding with a summary of key results (Section~\ref{sec:summ}).

\section{Supercritical forcing: basis and interpretation in eddying atmospheres}
\label{sec:why-relevant}

\subsection{Solsticial insolation}
\begin{figure}[t]
  \centering\noindent
  \includegraphics[width=0.5\textwidth, trim={0 0 0 0},clip]{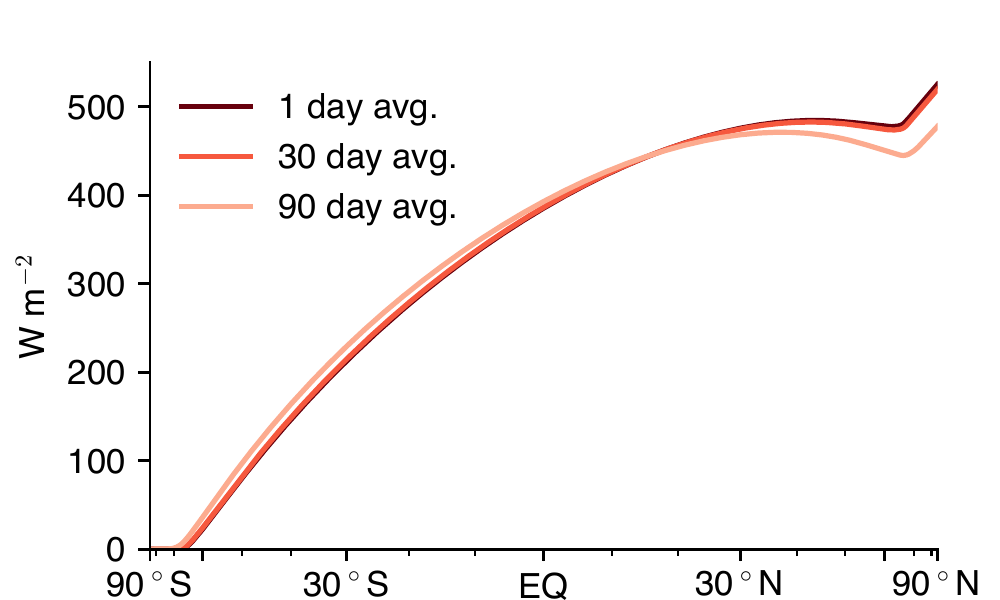}\\
  \caption{Insolation for averaging windows centered on northern summer solstice of 1, 30, and 90 days, in \wm/.  It is computed using the ``daily\_insolation'' function of the \texttt{climlab} package \citep{rose_climlab:_2018}, and is based on the methods of \citet{berger_insolation_1991}.}
  \label{fig:insol}
\end{figure}

Fig.~\ref{fig:insol} shows the diurnally averaged insolation distribution on the day of boreal summer solstice for Earth's present-day orbit (all results are equally applicable to austral summer).  Insolation is zero in the polar night region spanning the winter high latitudes.  Moving northward, it increases, reaching \(\sim\)386~\wm/ at the equator, but with steadily decreasing slope up to a local maximum of \(\sim\)485~\wm/ near 43\degr{}N.  From there it decreases modestly to a local minimum of \(\sim\)478~\wm/ near 62\degr{}N and finally increases monotonically from there to its global maximum of \(\sim\)525~\wm/ at the north pole.  Fig.~\ref{fig:insol} also shows insolation for longer averaging periods of 30 and 90 days centered on northern summer solstice.  Differences across the three averaging periods are modest.

\subsection{Conceptual basis of supercritical forcing extent}
If RCE prevailed at each latitude, then large-scale meridional and vertical velocities would vanish.  The large-scale zonal velocity field would be in gradient balance (\ie/ thermal wind balance but also including the nonlinear metric term) with the temperature field that is determined by the interactions between radiative and convective processes at each latitude.  But this exhibits physically untenable features, most obviously at the equator where the Coriolis parameter vanishes: no gradient-balanced solution is attainable with a nonzero cross-equatorial insolation gradient (which occurs at all times other than equinox; LH88).

Away from the equator in the summer hemisphere where RCE temperature increases moving poleward, gradient balance yields upper-tropospheric easterlies (assuming zonal wind is small at the surface due to drag) that draw angular momentum below its local planetary value.
If sufficiently strong, these easterlies can cause the RCE angular momentum field, denoted \(\Mrce\), to be increasing poleward, thereby changing the sign of the \(\Mrce\) meridional gradient and thus of the RCE absolute vorticity, denoted \(\etarce\) \citep{plumb_response_1992}.  Symbolically, this implies \(f\etarce<0\) \citep{emanuel_thermally_1995}, where \(f\equiv2\Omega\sinlat\) is the planetary vorticity (\ie/ the Coriolis parameter) with \(\Omega\) planetary rotation rate and \(\lat\) latitude.  That is impossible to sustain for multiple reasons \citep[see][for details]{adam_global_2009,hill_axisymmetric_2019}: it implies local extrema in \(\Mrce\), which cannot be sustained in the presence of nonzero viscosity; it is the sufficient condition for symmetric instability; and, near the tropopause where vertical velocity vanishes, a change in sign would require the absolute vorticity to pass through a fixed point (\ie/ where \(\pdsl{\etarce}{t}=0\)) that occurs at \(\etarce=0\) in the vorticity equation.  A large-scale circulation must emerge spanning at minimum all such latitudes, which are referred to as supercritical.  Equivalently, where \(\etarce=0\) in the summer hemisphere constitutes the minimal extent of the large-scale circulation in that hemisphere.\footnote{A latitude is also supercritical if \(\Mrce>\Omega a^2\) or \(\Mrce<0\) \citep{held_nonlinear_1980}.  But in the summer hemisphere at least for Earth, the \(\etarce=0\) point sits poleward of these conditions, save perhaps for just after spring equinox when the \(\Mrce=\Omega a^2\) point can be farther \citep[\cf/ Figs.~3 and 4 of][]{hill_axisymmetric_2019}.  Henceforth we take the summer-hemisphere supercritical forcing extent as identical to where \(\etarce=0\).}

\subsection{Supercritical forcing in eddying atmospheres}
Supercritical forcing extent has not figured centrally in theories for Earth's solsticial Hadley cell ascending edge for reasons that seem plausible in passing but that falter under scrutiny.

First is the notion that supercriticality is meaningful in axisymmetric atmospheres only and is \emph{in principle} inapplicable to macroturbulent atmospheres.  One can see how this would emerge.  Supercriticality (though not referred to as such) was popularized by \citet{held_nonlinear_1980} as an intermediate step in developing their highly influential axisymmetric, angular-momentum-conserving model for the annual-mean Hadley cells.  For solstice, the \(f\etarce<0\) facet was presented by \citet{plumb_response_1992} also in a purely axisymmetric context \citep[though soon extended to moist, zonally varying contexts by][]{emanuel_thermally_1995}.  Moreover, the marginally critical state of \(\etarce=0\) corresponds to uniform \(\Mrce\), which, with its homogeneous angular momentum distribution, might sound like a description of the axisymmetric (and nearly inviscid) angular-momentum-conserving model.

But the angular momentum that is spatially homogeneous in the angular-momentum-conserving model is that of the dynamically equilibrated state, \(M\), and crucially \(\Mrce\neq M\).  By definition, the latitude-by-latitude RCE state is one in which there is no large-scale circulation, zonally symmetric or otherwise.  Irrespective of whether the Hadley cells in the dynamically equilibrated state end up perfectly homogenizing angular momentum, or are totally controlled by eddies, or (most likely) something in between, latitude-by-latitude RCE cannot be sustained over any latitude that is supercritically forced.  Therefore, at least in the narrow sense regarding the minimal extent of a large-scale circulation of some kind, supercritical forcing extent is meaningful in all rotating atmospheres.

This leads to a second concern: whether \emph{in practice} the supercritical forcing extent usefully predicts, much more specifically, the location of the Hadley cell ascending edge.  In simulations for which supercritical extent has been explicitly computed, the ascending edge latitude sits poleward of the \(\etarce=0\) latitude \citep{faulk_effects_2017,hill_axisymmetric_2019,singh_limits_2019}.\footnote{Earth's extratropics, which are nominally subcritical by this definition throughout the annual cycle, obviously are not in a state of latitude-by-latitude RCE.  There, the hypothetical RCE state is unstable in other ways, of most relevance baroclinically.  Such baroclinic instability --- and with it an extratropical dynamical regime --- could in principle extend into the supercritically forced region, pushing the solsticial Hadley cell ascending latitude equatorward thereof \citep[much as it limits the Hadley descending, poleward edges, \cf/][]{held_general_2000,korty_extent_2008,kang_expansion_2012}.}  As such, to be a useful predictor, the supercritical forcing extent must scale proportionally with the actual ascending cell edge latitude.  As Section~\ref{sec:sims} will demonstrate --- albeit empirically --- this does in fact hold in a diverse range of idealized GCM simulations.

\section{Latitude-by-latitude RCE under solsticial forcing}
\label{sec:rce-sims}

\subsection{Numerical simulations}
We use the \texttt{climlab} single-column model \citep{rose_climlab:_2018} to simulate solsticial latitude-by-latitude RCE.  Each single-column simulation is forced with insolation corresponding to present-day, boreal summer solstice at a specified latitude, with the chosen latitudes in 1\degr{} increments spanning from equator to the pole in the summer hemisphere and from the equator to 55\degr{} in the winter hemisphere.
Apart from using solsticial rather than annual-mean insolation, the setup is identical to that of \citet{hill_axisymmetric_2020}, to which readers are referred for more details.

\begin{figure}[t]
  \centering\noindent
  \includegraphics[width=0.5\textwidth]{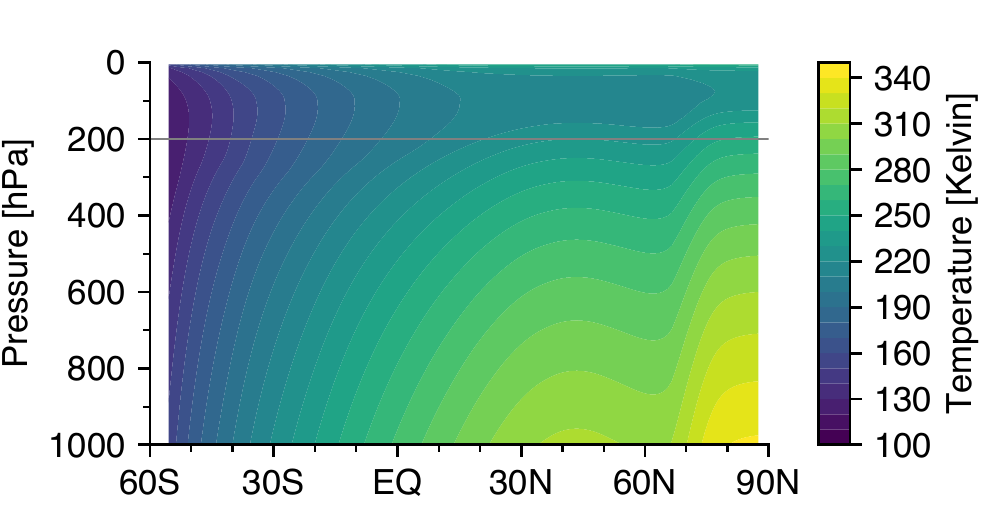}\\
  \caption{Temperature as a function of latitude and pressure from the solsticial RCE simulation, as indicated in the colorbar.  The gray line at 200~hPa indicates the level at which temperature is used to compute the gradient-balanced wind.}
  \label{fig:climlab-temp}
\end{figure}

Time-averaged fields from the single-column simulations are concatenated together in latitude to yield latitude-pressure distributions of each field.  Fig.~\ref{fig:climlab-temp} shows the resulting temperature field, \(T\).  From the temperature distribution, zonal wind at each level is inferred by assuming gradient wind balance and integrating the gradient balance expression from the surface where \(u\approx0\) is assumed to the given level:
\begin{align}
  \label{eq:grad-wind-p-coords}
  u(p,&\lat)=\nonumber\\
  &\Omega a\coslat\left[\sqrt{1 - \frac{1}{\coslat\sinlat}\frac{R_\mr{d}}{\Omega^2a^2}\ln\left(\frac{p_s}{p}\right)\pdlat{\hat T}} - 1\right],
\end{align}
where \(\hat T\) is the log-pressure-weighted average temperature from the surface pressure \(\ps=1000\)~hPa to the given pressure \(p\), and \(R_\mr{d}\) is the dry air gas constant.  We restrict attention to values at a specified tropopause pressure of 200~hPa.  Results are qualitatively insensitive to reasonable variations in the tropopause treatment, an issue explored at length by \citet{hill_axisymmetric_2020}.

From this zonal wind field, the angular momentum and absolute vorticity fields are subsequently calculated.  Specifically, angular momentum is
\begin{equation}
  \label{eq:ang-mom}
  M=a\coslat(\Omega a\coslat+u),
\end{equation}
and absolute vorticity is proportional to the meridional derivative of absolute angular momentum:
\begin{equation}
  \label{eq:abs-vort}
  \eta=\dfrac{-1}{a^2\coslat}\pd{M}{\lat}=f+\zeta,
\end{equation}
where \(\zeta=-(a\coslat)^{-1}\pdsl{(u\coslat)}{\lat}\) is the relative vorticity.

\begin{figure}[t]
  \centering\noindent
  \includegraphics[width=0.48\textwidth,angle=0]{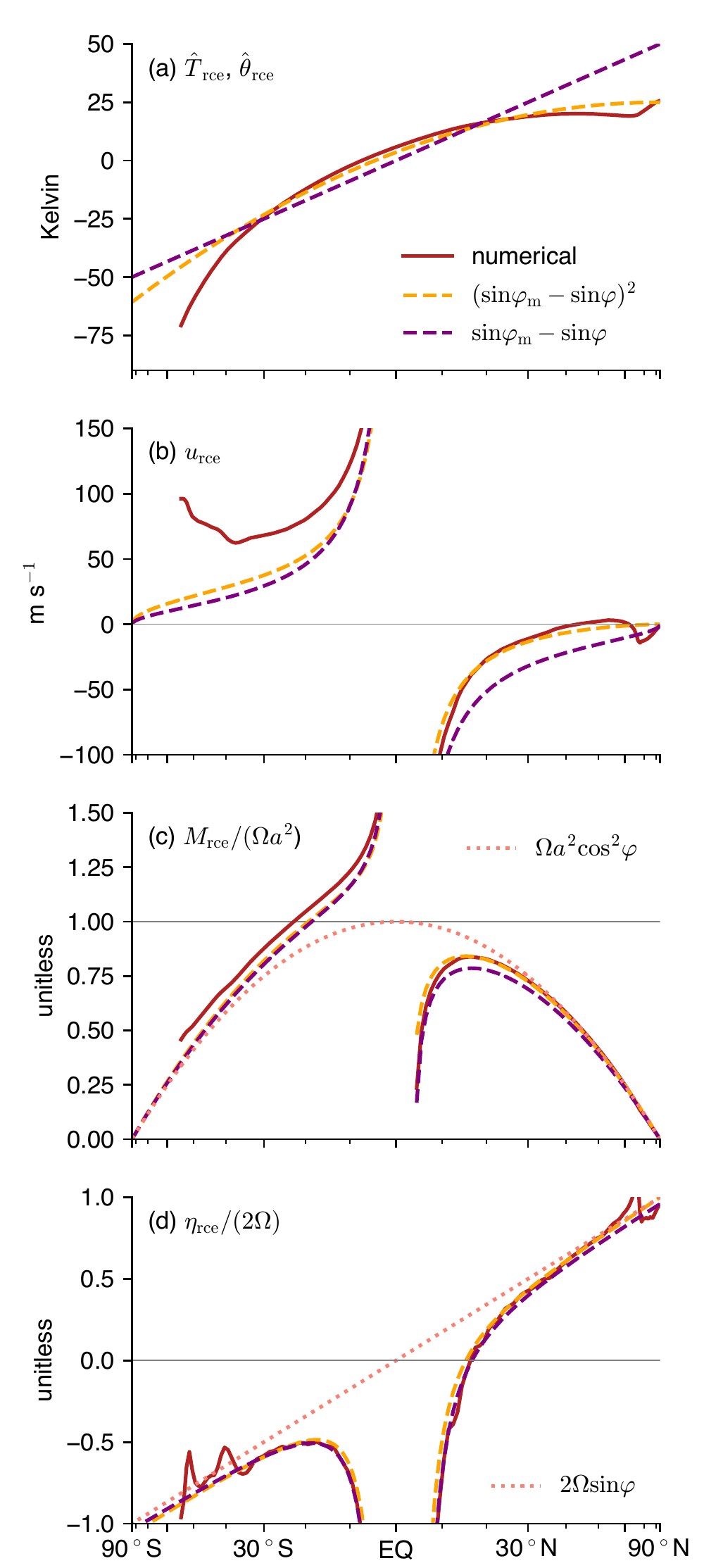}\\
  \caption{In solid red, results from numerical simulations of latitude-by-latitude radiative-convective equilibrium, compared to approximations thereto as dashed curves.  Dashed yellow corresponds to the analytical forcing profile given by Eq.~(\ref{eq:lh88-forcing}), and dashed purple corresponds to the further simplified forcing that is linear rather than quadratic in \(\sin\maxlat-\sin\lat\).  Panels, from top to bottom: (a) vertically averaged temperature or potential temperature, shown as deviation from 45\degr{}S-45\degr{}N mean; (b) gradient-balanced zonal wind at the tropopause; (c) absolute angular momentum at the tropopause; and (d) absolute vorticity at the tropopause.}
  \label{fig:climlab-vs-lh88}
\end{figure}

The solid curves in Fig.~\ref{fig:climlab-vs-lh88} show the simulated meridional profiles of temperature averaged from the surface to 200~hPa and of the inferred 200-hPa zonal wind, absolute angular momentum, and absolute vorticity.  The depth-averaged temperature field (shown as a deviation from its 45\degr{}S-45\degr{}N mean)  retains the extrema locations of the insolation and varies meridionally by roughly 25~K from the equator to the summer pole and 75~K from the equator to the region of polar night.  The inferred gradient wind is westerly throughout the winter hemisphere and asymptotes toward infinity approaching the equator; it is undefined in a narrow range of the summer hemisphere near the equator, poleward of which very strong easterlies gradually weaken, turning to weak westerlies around 40\degr{}N.  This zonal wind field causes the angular momentum field to deviate sharply from its planetary value (overlain in panel c).  Angular momentum is undefined from the equator to \(\sim\)5\degr{}N and increases to a local maximum near \(\sim\)15\degr{}N, poleward of which it tends toward the planetary value as \(u\) weakens and the distance from the rotation axis diminishes.  The absolute vorticity field changes sign at the angular momentum maximum \(\sim\)15\degr{}N, and this constitutes the poleward extent of supercritical forcing in the summer hemisphere.

\subsection{Analytical approximation}

We approximate the numerically simulated RCE state using the equilibrium temperature profile originally presented by LH88.  It is specified in terms of potential temperature averaged at each latitude over the fixed depth \(H\) of a Boussinesq atmosphere and may be written
\begin{equation}
  \label{eq:lh88-forcing}
  \dfrac{\hat\theta\rce}{\theta_0}=1+\dfrac{\deltah}{3}\left[1-3(\sin\maxlat-\sinlat)^2\right],
\end{equation}
where \(\theta\) is potential temperature, the hat denotes a depth average, \(\theta_0\) is the Boussinesq reference potential temperature, \(\hat\theta\) maximizes at the latitude \(\maxlat\), and \(\deltah\) is a parameter controlling (in conjunction with \(\maxlat\)) the fractional variations in \(\hat\theta\rce\) with latitude.  The ``rce'' subscript emphasizes that we are treating \eqref{eq:lh88-forcing} as an approximation to the hypothetical latitude-by-latitude RCE state that would occur absent a large-scale circulation.

The Boussinesq expression for gradient-balanced zonal wind at height \(H\) is nearly identical to (\ref{eq:grad-wind-p-coords}), with \(R_\mr{d}\ln(p/\ps)\pdsl{\hat T}{\lat}\) replaced by \((gH/\theta_0)\pdsl{\hat\theta}{\lat}\):
\begin{equation}
  \label{eq:u-rce}
  u=\Omega a\coslat\left[\sqrt{1-\frac{1}{\coslat\sinlat}\frac{gH}{\Omega^2a^2\theta_0}\pdlat{\hat{\theta}}}-1\right],
\end{equation}
where \(g\) is gravity and the surface zonal wind has been assumed negligible due to surface friction.  We use (\ref{eq:lh88-forcing}) as \(\hat\theta\) in this expression to find \(\urce\); \(\Mrce\) and \(\etarce\) then follow using \eqref{eq:ang-mom} and \eqref{eq:abs-vort} (the corresponding analytical expressions are shown in the next section).

With \(g\), \(\Omega\), and \(a\) set to appropriate Earth values, there are still four free parameters between (\ref{eq:lh88-forcing}) and (\ref{eq:u-rce}), namely \(H\), \(\theta_0\), \(\maxlat\), and \(\deltah\) --- enough to potentially over-tune to the numerically simulated RCE fields of interest.  Appendix~A details our procedure for choosing these values; in short, we choose conventional values of \(H=10\)~km and \(\theta_0=300\)~K and then perform a two-dimensional parameter sweep over \(\deltah\) and \(\maxlat\) values to find best fits to the simulated RCE temperature field over 45\degr{}S-45\degr{}N (rather than directly for the \(\etarce=0\) point of ultimate interest).  Fortunately, provided \(\maxlat\gtrsim30^\circ\), the product \(\deltah\sin\maxlat\) --- which will figure centrally in our scaling below --- is nearly constant, provided that for each \(\maxlat\) one sets \(\deltah\) to its best-fit value for that \(\maxlat\).

The resulting \(\hat\theta\rce\), \(\urce\), \(\Mrce\), and \(\etarce\) fields with \(\maxlat=90\)\degr{}N and \(\deltah=1/15\) are overlain in Fig.~\ref{fig:climlab-vs-lh88} as dashed orange curves.  In short, the LH88 approximation captures the numerically simulated RCE state well throughout most of the domain of relevance to the Hadley cells.  In more detail, the numerically simulated depth-averaged temperature field has greater meridional curvature than the LH88 solution in the extratropics, but at lower latitudes of more relevance to the Hadley cells the two are nearly coincident.  The same largely holds for the zonal wind, though it begins to deviate substantially (\(\gtrsim20\)\ms/) from the LH88 solution by the southern subtropics and deviates further poleward thereof.  The effect of this is weaker, however, on the angular momentum and absolute vorticity fields.  In the summer hemisphere the absolute vorticity field is very accurately captured by the LH88 field deep into the extratropics --- including the zero crossing near \(\sim\)15\degr{}N that constitutes the poleward edge of the supercritical forcing extent.

\section{Analytical expression for solsticial supercritical forcing extent}
\label{sec:theory-summer}

Inserting \eqref{eq:lh88-forcing} into \eqref{eq:u-rce} yields the gradient-balanced zonal wind under LH88 forcing,
\begin{equation}
  \label{eq:u-rce-lh88}
  \urce=\Omega a\coslat\left[\sqrt{1+2\Roth\left(\frac{1}{\sin\maxlat}-\frac{1}{\sinlat}\right)}-1\right],
\end{equation}
where
\begin{equation}
  \label{eq:therm-ro}
  \Roth\equiv\dfrac{gH}{\Omega^2a^2}\deltah\sin\maxlat
\end{equation}
is the thermal Rossby number.  Equivalently \({\Roth=\Bu\deltah\sin\maxlat}\), where \(\Bu\equiv gH/(\Omega a)^2\) is the planetary Burger number.  Our inclusion of \(\sin\maxlat\) in the thermal Rossby number is nonstandard and makes  (\ref{eq:therm-ro}) relevant to solsticial seasons only (since \(\sin\maxlat=0\) for the equinoctial seasons and the annual mean).  It is motivated by Appendix~A, which shows that different fits of the LH88 forcing to the solsticial RCE state largely collapse onto a single value of \(\deltah\sin\maxlat\) (for \(\maxlat\) values outside the tropics, as is appropriate).

Using \eqref{eq:u-rce-lh88} in \eqref{eq:ang-mom} then yields the corresponding absolute angular momentum field,
\begin{equation}
  \label{eq:ang-mom-lh88}
  \Mrce=\Omega a^2\cos^2\lat\sqrt{1+2\Roth\left(\frac{1}{\sin\maxlat}-\frac{1}{\sinlat}\right)},
\end{equation}
and similarly using \eqref{eq:u-rce-lh88} in \eqref{eq:abs-vort} yields the corresponding absolute vorticity field:
\begin{align}
  \label{eq:abs-vort-lh88}
  \etarce = 2\Omega\sinlat\sqrt{1+2\Roth\left(\frac{1}{\sin\maxlat}-\frac{1}{\sinlat}\right)}\nonumber\\
  \times\left[1-\frac{1}{2}\frac{\cos^2\lat}{\sin^3\lat}\frac{\Roth}{1+2\Roth\left(\frac{1}{\sin\maxlat}-\frac{1}{\sinlat}\right)}\right].
\end{align}

Eq.~\eqref{eq:abs-vort-lh88} comprises three terms multiplying one another.  The first is simply the local planetary vorticity, \(f\), which is irrelevant to the zero crossing within the summer hemisphere.  The second, the square-root term, amounts by \eqref{eq:ang-mom-lh88} to \(\Mrce/(\Omega a^2\cos^2\lat)\).  Its zero crossing corresponds to the latitude very near the equator where \({\Mrce=0}\).  Here \(\urce\) is strongly negative, and it becomes less so moving toward \(\maxlat\) such that \(\Mrce\) increases, and thus \(f\etarce<0\), over some span poleward of this point.  Therefore, the actual \(\etarce=0\) point in the summer hemisphere always sits poleward of the \({\Mrce=0}\) point \citep[see Fig.~3a of][]{hill_axisymmetric_2019} and depends on the third term in \eqref{eq:abs-vort-lh88}, \ie/ everything within the large square brackets.

Without approximation, the third term vanishes at the latitude \(\latcrit\) satisfying
\begin{equation}
  \label{eq:eta-term3-full}
  \left(1+2\dfrac{\Roth}{\sin\maxlat}\right)\sin^3\latcrit-\dfrac{3}{2}\Roth\sin^2\latcrit-\dfrac{1}{2}\Roth=0.
\end{equation}
An exact solution to this third-order polynomial in \(\sin\latcrit\) can be found using the cubic formula, but its form (not shown) is too complicated to draw physical insights from.  We therefore pursue an approximate solution as follows.  If we assume \({0<\Roth\ll\sin\maxlat\leq1}\) and  \({0<\latcrit\ll\sin\maxlat\leq1}\), then \(\latcrit\approx\sin\latcrit\) and to leading order (\ref{eq:eta-term3-full}) becomes
\begin{equation}
  \label{eq:eta-term3-approx1}
  \latcrit^3-\dfrac{3}{2}\Roth\latcrit^2-\dfrac{1}{2}\Roth=0.
\end{equation}
This is only meaningful if \({\Roth\ll\latcrit}\), since \({\Roth\sim\latcrit}\) would lead to a self-contradictory balance between terms of order \(\Roth^3\) with a term of order \(\Roth\) (or equivalently \(\latcrit^3\) with \(\latcrit\)).\footnote{A third mathematically possible case, \(0<\latcrit\ll\Roth\ll\sin\maxlat\), yields a physically nonsensical result.}  Thus, assuming \(0<\Roth\ll\latcrit\ll\sin\maxlat\), the approximate solution to \eqref{eq:eta-term3-approx1} is simply
\begin{equation}
  \label{eq:eta-term3-approx2}
  \latcrit=\left(\dfrac{\Roth}{2}\right)^{1/3}.
\end{equation}
According to \eqref{eq:eta-term3-approx2}, the solsticial Hadley ascending edge latitude varies with the thermal Rossby number to the one-third power.

As shown in Appendix~B, a \(\Roth^{1/3}\) scaling for the supercritical forcing extent also emerges for any \({\hat\theta\rce\propto(\sin\maxlat-\sinlat)^n}\) with integer \({n\geq1}\).  That more general solution is
\begin{equation}
  \label{eq:poly-sol}
  \latcrit=\left(\dfrac{n\Roth}{4}\right)^{1/3}.
\end{equation}
This includes the \({n=1}\) case in which the forcing is simply linear in \(\sin\lat\).  This can be seen from the overlain dashed purple curves in Fig.~\ref{fig:climlab-vs-lh88}, which are the \(\hat\theta\rce\), \(\urce\), \(\Mrce\), and \(\etarce\) fields computed with \({n=1}\), \(\maxlat=90\)\degr{}N, and \({\deltah=2/15}\), \ie/ twice the value used for the \({n=2}\) case, such that \(n\Roth\) is the same between them.  Though certainly less accurate than the \({n=2}\) approximation overall, the \({n=1}\) case captures the numerically simulated RCE behavior in the tropics suitably.  We conclude that, with respect to the supercritical forcing extent, the extratropical wiggles and meridional curvature in the tropics of the solsticial insolation matter little compared to the overall increase moving toward the summer pole.

\begin{figure*}[t]
  \centering\noindent
  \includegraphics[width=\textwidth]{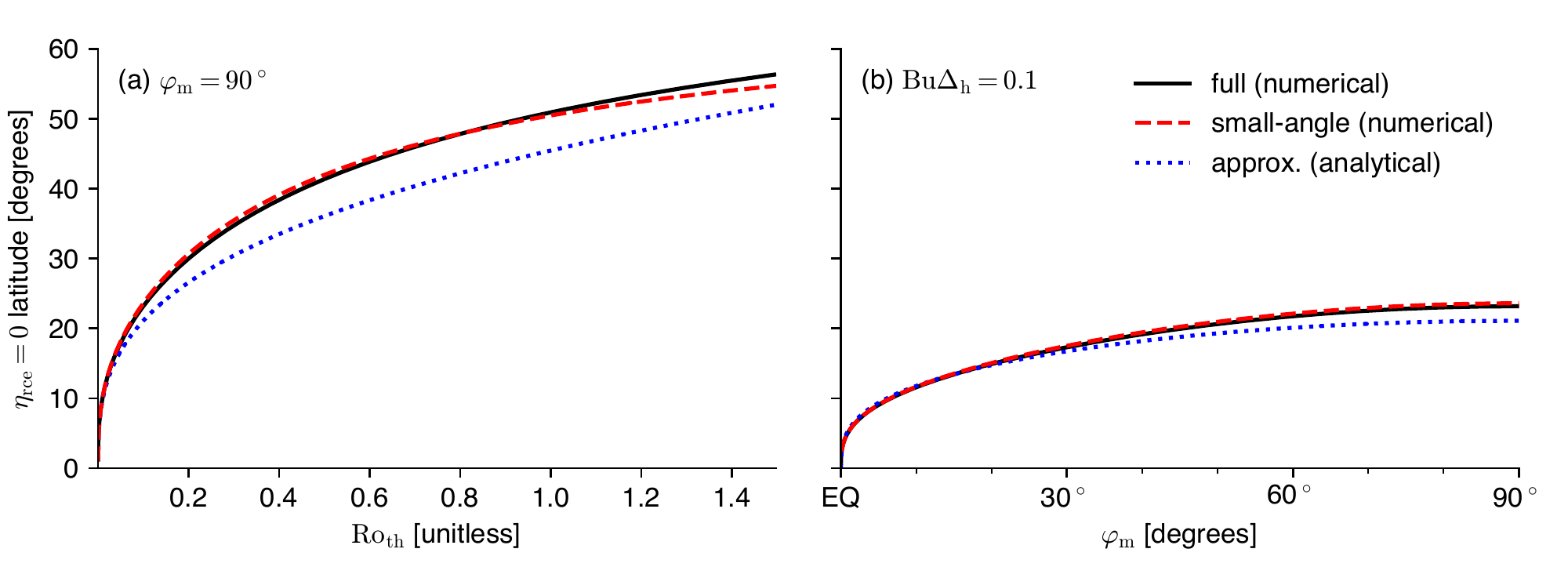}\\
  \caption{Supercritical forcing extent under the forcing given by (\ref{eq:lh88-forcing}) as a function of different parameters, with the full numerical solution, the small-angle numerical solution, and the analytical solution given by (\ref{eq:eta-term3-approx2}) as indicated in the legend in panel b.  In panel a, \(\maxlat=90^\circ\) and solutions are shown as a function of \(\Roth\).  In panel b, \(\Bu\deltah=0.1\) and solutions are shown as a function of \(\maxlat\).}
  \label{fig:etarce-phase-space}
\end{figure*}

Fig.~\ref{fig:etarce-phase-space}(a) shows the supercritical forcing extent, \ie/ where (\ref{eq:abs-vort-lh88}) vanishes, solved numerically, if \(\maxlat=90^\circ\) as \(\Roth\) is varied over \(0<\Roth<1.5\), and Fig.~\ref{fig:etarce-phase-space}(b) shows the same but with \(\Bu\deltah=0.1\) as \(\maxlat\) is varied from equator to pole.  Fig.~\ref{fig:etarce-phase-space} also shows numerical solutions for the small-angle approximation \eqref{eq:eta-term3-approx1} and the analytical expression (\ref{eq:eta-term3-approx2}).  For the given \(\maxlat=90^\circ\) (panel a), the true zero crossing and the approximation thereto move poleward monotonically with \(\Roth\).  The approximation \eqref{eq:eta-term3-approx2} captures the exact expression reasonably well even for \({\Roth\sim1}\), though it is consistently equatorward of the exact value by a modest degree.  Similarly, for a reasonably Earth-like \({\Bu\deltah\sim0.1}\), the zero crossing moves poleward most rapidly as \(\maxlat\) moves off the equator by a few degrees and increases more gradually poleward thereof (panel b).  In the small-angle approximation, for example, the maximum value of 23.6\degr{} occurs for \(\maxlat=90^\circ\), but it is displaced only 2\degr{} equatorward thereof for \(\maxlat\) moved all the way to 55\degr{}N.  The approximate solution again is accurate though biased slightly equatorward for large \(\maxlat\).


Finally, as noted above the actual solsticial Hadley cell ascending edge, \(\ascentlat\) tends to be displaced poleward of \(\latcrit\) by a constant multiplicative factor.  But we do not have a theory for that factor, which furthermore will prove to vary across idealized GCMs in the simulations in the next section.  As such, from (\ref{eq:eta-term3-approx2}) we arrive at a scaling (rather than precise prediction) for \(\ascentlat\):
\begin{equation}
  \label{eq:ascent-scale}
  \ascentlat\propto\Roth^{1/3}.
\end{equation}

We deem noteworthy and worth future exploration that this scaling is essentially the same as that derived by \citet[][\cf/ their Eq.~56]{caballero_axisymmetric_2008} for the \emph{descending} edge in the winter hemisphere, despite seemingly unrelated sets of assumptions between the two studies.  Specifically, motivated by their numerical, axisymmetric simulations, \citet{caballero_axisymmetric_2008} assume that the Hadley cell zonal wind field conserves angular momentum from the equator to the winter-hemisphere descending edge and that the descending edge latitude is proportional to the ascending edge latitude \(\ascentlat\); they then use equal-area arguments to find a one-third power law scaling with the thermal Rossby number of the descending edge latitude (and implicitly of \(\ascentlat\)).  We make no assumptions about the Hadley cell zonal wind field that emerges, with our scaling for \(\ascentlat\) emerging instead as where the latitude-by-latitude RCE absolute vorticity vanishes in the summer hemisphere.

\section{Ascending edge latitude in idealized GCM simulations}
\label{sec:sims}

Among other things, (\ref{eq:ascent-scale}) implies \({\ascentlat\propto\Omega^{-2/3}}\).  Here we present evidence that this accurately characterizes the moist idealized GCM simulations originally presented by \citet[][hereafter F17]{faulk_effects_2017} and \citet[][hereafter S19]{singh_limits_2019} as well as newly performed simulations in an idealized dry GCM.  Each of the three idealized GCMs --- the idealized aquaplanet of \citet{frierson_gray-radiation_2006} for F17, the version thereof as modified by \citet{ogorman_hydrological_2008} for S19, and the dry idealized GCM of \citet{schneider_tropopause_2004} --- are widely used and documented in the literature.  As such, we leave details of model formulation in Appendix~C and describe here only the salient properties specific to the simulation sets used here.

\subsection{Description of simulations}
For the F17 simulations, insolation follows the present-day Earth annual cycle, diurnally averaged, using a 360-day calendar.  Across the simulations, planetary rotation rate is varied by factors of two from 4\(\times\) to 1/32\(\times\Omega_\mr{E}\), where \(\Omega_\mr{E}\) is Earth's value, as well as one with 1/6\(\times\Omega_\mr{E}\), with all other planetary parameters taking their standard Earth values.  The simulations are run at T42 horizontal spectral resolution, with 25 levels unevenly spaced in the \(\sigma\) vertical coordinate, and for ten 360-day years.  Results are averaged over the 30 days centered on northern solstice across the last eight years.\footnote{This deviates from the procedure of F17, who vary their 40-day solsticial averaging window across simulations based on the seasonal timing of the ITCZ poleward migration into the summer hemisphere.  Results are qualitatively insensitive to this difference (not shown).}  Three additional simulations, at 1, 1/8, and \(1/32\times\Omega_\mr{E}\), are forced with time-invariant solsticial rather than seasonally varying insolation, and we present averages over the final eight years of these 10-year integrations.

The simulations of S19 were run at T42 spectral horizontal resolution with 30 unevenly spaced \(\sigma\)-levels.  Rather than seasonally varying insolation, these simulations are forced at all times by the diurnally averaged insolation occurring at present-day northern solstice.  Planetary rotation rate is varied across the simulations, one each for 8, 4, 3, 2, 3/2, 1, 3/4, 2/3, 1/2, 1/4, and 1/8\(\times\Omega_\mr{E}\).  The simulations span \(6\times360=2,160\)~days, and results are averaged over the final 720~days.

In the idealized dry GCM, radiative transfer is approximated by Newtonian cooling toward a prescribed equilibrium temperature profile, which thereby defines the hypothetical latitude-by-latitude RCE temperature field.  As such, we set its meridional structure to be (\ref{eq:lh88-forcing}), with \({\theta_0=300}\)~K, \(\maxlat=90^\circ\) and \(\deltah=1/15\), the same values as used in Section~\ref{sec:rce-sims}.  Simulations are performed with planetary rotation rates of 2, 1, and 1/4\(\times\Omega_\mr{E}\) with \(\deltah=1/15\).  One additional sensitivity test is performed at Earth's rotation rate with \(\deltah=1/6\) as in LH88 (though the largest \(\maxlat\) used by that study was 8\degr{}).  All simulations ran for 1440 days, with averages taken over the final 360 days.  We refer to these as the LH88-forced simulations.

For all simulations, we compute the Hadley cell ascending edge latitude using the definition of S19, as described in Appendix~C.  We diagnose \(\Roth\) for each simulation using the appropriate value of \(\Omega\), standard Earth values of \(a\) and \(g\), the \(\sin\maxlat=1\) and \(\deltah=1/15\) best-fit values inferred from the latitude-by-latitude RCE simulations.  For the approximate RCE tropospheric depth \(H\), we infer it to be \(\sim\)10~km for the F17 and S19 simulations based on the explicit latitude-by-latitude RCE simulation performed by S19.\footnote{Specifically, from Fig.~5 of S19, over the summer-hemisphere latitudes relevant to supercriticality the troposphere-average temperature (\(\hat T\)) is \(\sim\)275~K, and the ratio of the tropopause and surface pressures (\(p_\mr{t}/\ps\)) is \(\sim\)0.35.  Ignoring virtual effects, the hypsometric equation then yields a tropopause height of \(H=(R_\mr{d}/g)\hat T\ln(\ps/p_\mr{t})\approx10\)~km.}  For the LH88-forced simulations, we infer \(H\) directly from the imposed equilibrium temperature field, yielding \(\sim\)7~km (not shown).

\subsection{Simulation results}
\begin{figure*}[t]
  \centering\noindent
  \includegraphics[width=\textwidth]{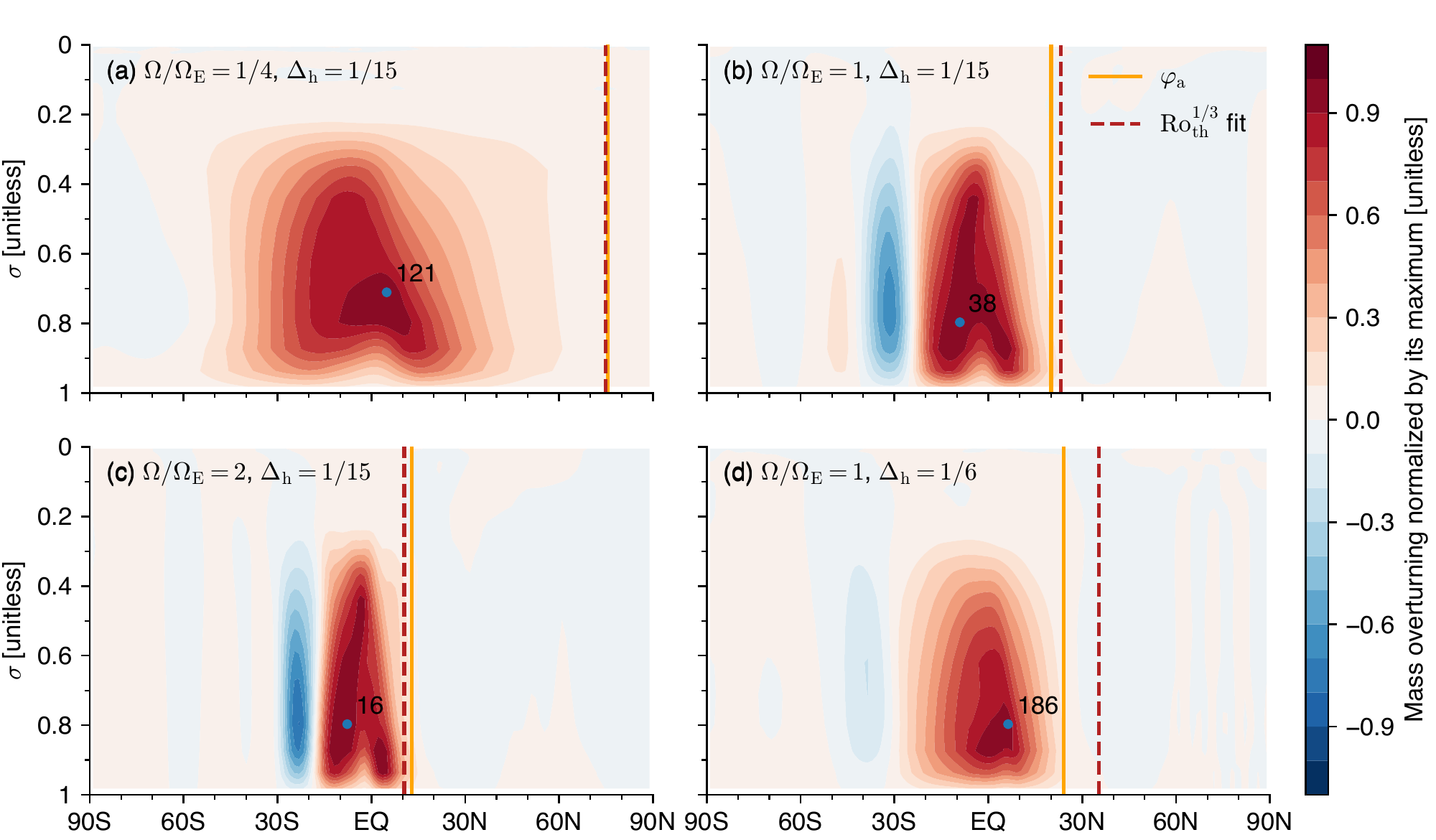}\\
  \caption{Mass overturning streamfunction normalized by its maximum value in each of the LH88-forced simulations.  Each panel corresponds to the simulation as labeled in the panel's top left corner, where \(\Omega_\mr{E}\) is Earth's rotation rate.  The blue dot indicates the solsticial cell maximum in the free troposphere, and the adjacent number indicates the mass overturning strength there, in \(10^9\)~kg~s\inv{}.  The vertical solid orange line in each panel is the simulated \(\ascentlat\) computed using (\ref{eq:edge-def}).  The vertical dashed red line is the approximation thereto from a linear best fit in \(\Roth^{1/3}\) across the three simulations with \(\deltah=1/15\).}
  \label{fig:lh88-streamfuncs}
\end{figure*}

Fig.~\ref{fig:lh88-streamfuncs} shows the mass overturning streamfunctions from the four LH88-forced simulations, each normalized by the solsticial Hadley cell's overall maximum overturning rate occurring at the cell center.  This facilitates comparison of the cell spatial structures across simulations in the face of large variations in strength, over an order of magnitude between the \({2\times\Omega_\mr{E}}\) case and the \({1\times\Omega_\mr{E}}\), \(\deltah=1/6\) case.  In the three \(\deltah=1/15\) cases, the weakness of the cross equatorial forcing gradient results in an equatorial jump of near-surface streamlines out of the boundary layer \citep[\cf/][]{pauluis_boundary_2004}.

Comparing to the streamfunctions of F17 (their Fig.~3 and 4) and S19 (his Fig.~3), at Earth's rotation rate there are differences in detail, but to first order the simulated cells are similar.  At 1/4\(\times\) Earth's rotation rate, there is more heterogeneity across the simulation sets, with the F17 cell extending the least far poleward and the S19 cell extending the farthest poleward.  Across all the simulations for each model the cross-equatorial Hadley cell grows as the planetary rotation rate decreases (as was shown by F17 and S19 and as expected for the LH88-forced simulations).

\begin{figure}[t]
  \centering\noindent
  \includegraphics[width=0.5\textwidth]{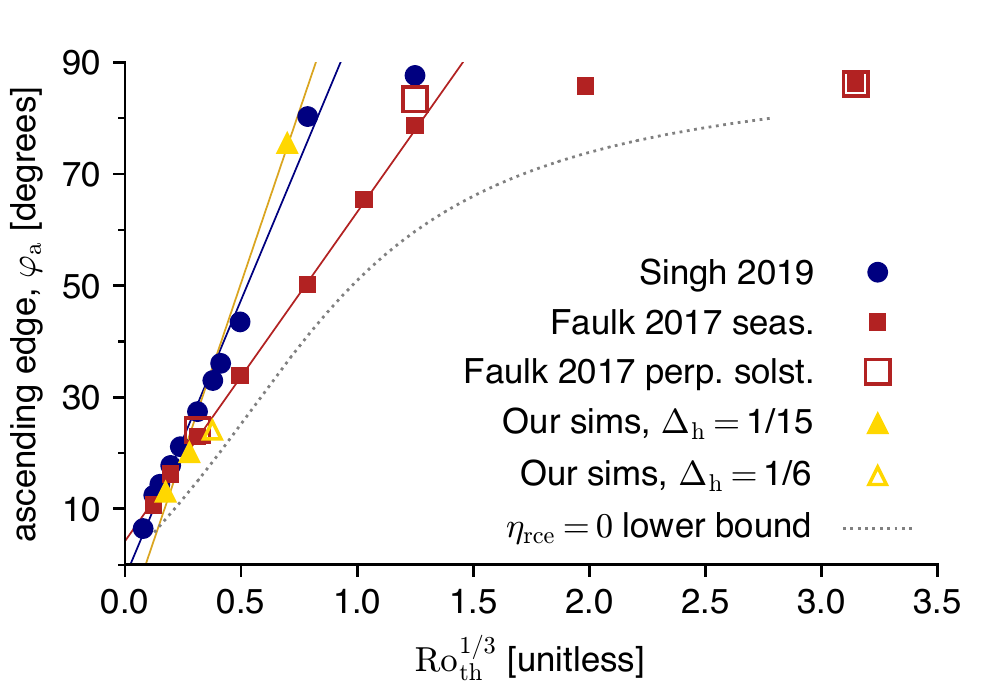}
  \caption{Cross-equatorial Hadley cell edge in the summer hemisphere in idealized aquaplanet simulations of \citet{faulk_effects_2017}, \citet{singh_limits_2019} and in the idealized dry simulations of the present study as a function of the thermal Rossby number to the one-third power, each signified by different symbols as indicated in the legend.  The solid lines show the linear best fit to \(\ascentlat\) as a function of \(\Roth^{1/3}\) for the given simulation set, restricting to \(\Roth<1\), with red, blue, and yellow for the S19, F17, and the \(\deltah=1/15\) dry simulations respectively.  The dotted gray curve is the numerical solution to (\ref{eq:eta-term3-full}).}
  \label{fig:sims-scaling}
\end{figure}

Fig.~\ref{fig:sims-scaling} shows the ascending edge latitude in each simulation as a function of \(\Roth^{1/3}\).  Plotted in this way, simulations that fall on a straight line, whatever their slope, scale with \(\Roth^{1/3}\) as (\ref{eq:ascent-scale}) predicts.  Overlain solid lines correspond to the linear best fit for each of the three simulation sets, restricted to simulations with \(\Roth<1\) where the small-angle and small-\(\Roth\) assumptions are plausible (for the LH88 simulations, this also does not include the outlier \(\deltah=1/6\) case, for reasons discussed below).  A \(\Roth^{1/3}\) scaling aptly characterizes each of the simulation sets in the relevant regime --- there is only moderate scatter for each simulation set about its linear best fit.  This includes simulations with \({\Roth\sim1}\), despite the scaling assuming \(\Roth\ll1\).

The slopes of the linear best fits in \((\Roth^{1/3}, \ascentlat)\) space for the F17, S19, and LH88-forced simulations are approximately 1.0, 1.7, and 2.1, respectively (with dimensions radians per \(\Roth^{1/3}\)).  The corresponding \(y\)-intercepts (in degrees latitude rather than radians) are approximately 4, -2, and -11, respectively.  Our scaling (\ref{eq:ascent-scale}) is agnostic to the slope but would predict a \(y\)-intercept of zero.  The proportionality constants span from nearly the lower bound of unity (F17 cases) to a little more than twice that (LH88-forced cases).  By eye from Fig.~\ref{fig:sims-scaling} and given the limited number of simulations and ambiguities in the estimate of \(H\), it is not clear how seriously the differences in the \(y\)-intercepts from zero should be taken.

Unfilled squares in Fig.~\ref{fig:sims-scaling} show \(\ascentlat\) in the three F17 perpetual solstice cases.  The ascending edge at \(1/32\times\Omega_\mr{E}\) is nearly identical for either insolation treatment, but in the \(1\times\) and \(1/8\times\Omega_\mr{E}\) time-invariant forced cases \(\ascentlat\) is a few degrees poleward from that of corresponding seasonally varying case.  This difference is not large, and the perpetual solstice F17 \(\ascentlat\) values still sit equatorward of the corresponding S19 ones.  We lack an explanation for this difference between the perpetual solstice simulations of F17 and S19, which is somewhat surprising given seemingly modest differences in model formulation.  The slope and \(y\)-intercept of the linear fit for the F17 \(1\times\) and \(1/8\times\Omega_\mr{E}\) time-invariant forced cases (1.1 radians per \(\Roth^{1/3}\) and 4\degr{} respectively) are also nearly identical to the annual-cycle counterparts.

We have also computed best fit power-law exponents by a standard least-squares fit to each simulation set in (\(\log_{10}\Roth, \log_{10}\ascentlat\)) space, again restricting to \(\Roth<1\).  For the F17, S19, and \(\deltah=1/15\) LH88-forced sets, the best fit \(\Omega\) exponents are 0.28, 0.34, and 0.41, respectively --- all reasonably close to the 1/3 power predicted by (\ref{eq:ascent-scale}), and nearly identical to it, at 0.34, in their average.  The exponent inferred for the \(1\times\) and \(1/8\times\Omega_\mr{E}\) F17 perpetual-solstice simulations is 0.30, slightly closer to the 1/3 value than the 0.28 value from the F17 seasonal cycle simulations.  Again given the uncertainties, this small difference may or may not be physically meaningful.

Finally, the overlain dotted curve in Fig.~\ref{fig:sims-scaling} shows the numerical solution to the full expression for \(\latcrit\), (\ref{eq:eta-term3-full}), which does not assume small-\(\Roth\) or small-\(\lat\).  This lower bound qualitatively captures the leveling off of \(\ascentlat\) in the F17 and S19 simulations with large \(\Roth\) where the cells become nearly pole-to-pole.

\subsection{LH88-forced case with \(\deltah=1/6\)}
The unfilled triangle in Fig.~\ref{fig:sims-scaling} corresponds to the LH88-forced simulation at Earth's rotation rate in which \(\deltah=1/6\) rather than 1/15 as in the others (but still with \(\maxlat=90^\circ\)).  The 2.5\(\times\) increase in \(\deltah\) increases \(\Roth\) accordingly, and the ascending edge latitude does move poleward, but not enough to fall along the same scaling as the \(\deltah=1/15\) cases.  This suggests that modifying \(\deltah\) at a fixed rotation rate excites one or more mechanisms that influence \(\ascentlat\) that the supercritical forcing extent does not account for.  This could constitute an important limitation to our theory's applicability to \eg/ changes under global warming.  Adjudicating this would require additional simulations and analyses beyond the scope of the present study, but we do speculate on one potential candidate, namely influences of \(\deltah\) on zonally asymmetric eddy processes.

In the \(\deltah=1/6\) case, the northern subtropics to extratropics exhibit a very long-lasting wave-3 pattern that propagates westward but persists for hundreds of days (not shown).  The wave is very regular.  It spans meridionally over \(\sim\)20-60\degr{}N, and its three centers are located between 30 and 40\degr{}N.  By contrast, in the \(\deltah=1/15\) case, the summer hemisphere zonally asymmetric circulation outside of the Tropics is much more Earth-like, with most commonly a wave-4 structure, but with individual lows and highs growing, decaying, and moving relative to each other, while on average being advected by the mean easterlies (not shown).  Such qualitatively distinct extratropical circulations in the summer hemisphere could very well impart very different influences on the Hadley circulation.

\section{Relationship to slantwise convective neutrality constraint}
\label{sec:disc}

In a state of slantwise convective neutrality, streamlines, angular momentum contours, and saturation moist entropy isolines are all parallel.  By assuming this characterizes the solsticial Hadley circulation, S19 derives a diagnostic, for the ascending edge latitude, which we denote \(\lats\) and can be written
\begin{equation}
  \label{eq:singh-diag}
  \sin^3\lats\cos\lats=\dfrac{\Delta T}{2\Omega^2a^2}\left.\dfrac{\partial s_b}{\partial \lat}\right|_\lats,
\end{equation}
where \(s_\mr{b}\) is the boundary layer moist entropy, and \(\Delta T\) is the difference between the boundary layer temperature at the latitude \(\lats\) and the temperature at the equatorial tropopause.   This expression corresponds to the latitude \(\lats\) at which an angular momentum contour --- and with it a streamline --- that emanates from the boundary layer crosses the equator at the tropopause, thereby constituting the outermost streamline of the cross-equatorial Hadley cell, \ie/ the ascending edge \(\ascentlat\).

S19 shows that this diagnostic predicts \(\ascentlat\) with quantitative accuracy across his simulations.  We showed above that in those same simulations \(\ascentlat\propto\latcrit\sim\Roth^{1/3}\sim\Bu^{1/3}\), where replacing \(\Roth\) with \(\Bu\) is justified since all parameters other than \(\Omega\) are constant.  Assuming that the stratification in low latitudes will be nearly moist adiabatic, we can approximate the lapse rate as \(\Gamma=\gamma\Gamma_\mr{d}\), where \(\Gamma_\mr{d}\equiv g/c_p\) is the dry adiabatic lapse rate and \(\gamma\approx0.7\), analogous to the convective adjustment scheme in the idealized dry GCM above.  We further assume that the tropopause temperature is meridionally uniform \citep{hill_axisymmetric_2020}, such that \(\Delta T\) in (\ref{eq:singh-diag}) can be replaced with the surface-tropopause temperature drop in the local column.\footnote{In equating the tropopause depth in the \(\lats\) expression --- which corresponds to the dynamically equilibrated state --- with that in the \(\latcrit\) expression --- which corresponds to the latitude-by-latitude RCE state --- we are implicitly assuming that the emergence of the circulation doesn't substantially change this depth.}  In that case, we have \(gH\approx c_p\Delta T/\gamma\), such that the leading factor on the RHS of (\ref{eq:singh-diag}) becomes \(\gamma\Bu/(2c_p)\).  Separately, by definition \(s_\mr{b}\equiv c_p\ln\theta_\mr{eb}\), where \(\theta_\mr{eb}\) is the sub-cloud equivalent potential temperature.  In the small-angle limit and recalling \eqref{eq:eta-term3-approx2}, this yields
\begin{equation}
  \label{eq:singh-diag-small-angle}
  \left(\dfrac{\lats}{\latcrit}\right)^3=\dfrac{\gamma}{\deltah\sin\maxlat}\left.\dfrac{\partial\ln\theta_\mr{eb}}{\partial\lat}\right|_\lats.
\end{equation}

Since \(\ascentlat\approx\lats\) and \(\ascentlat\propto\latcrit\) in the S19 simulations, and since \(c_p\), \(\gamma\), and \(\deltah\) are all constants, it follows that the boundary layer moist entropy gradient at the cell edge is itself constant across the simulations:
\begin{equation}
  \label{eq:const-eb}
  \left.\dfrac{\partial\ln\theta_\mr{eb}}{\partial\lat}\right|_\ascentlat\sim\mathrm{constant}.
\end{equation}
S19 notes that in the small-angle limit \(\left.\partial_\lat s_b\right|_\ascentlat\) edge must be small \citep[and thus the cell edge sits near a local \(s_b\) maximum, \cf/][]{prive_monsoon_2007}, but this does not constrain it to be constant.  We deem this worthy of future study.
One potentially important distinction between the slantwise convective neutrality diagnostic and our supercriticality-based theory is that \(\deltah\sin\maxlat\) appears in the latter but not the former.  This is as it should be, since \(\deltah\sin\maxlat\) characterizes the RCE state which the supercriticality depends on, while the slantwise convective neutrality diagnostic is a statement about the dynamically equilibrated state.  In other words, \(\lats\propto\Bu^{1/3}\), whereas \(\latcrit\propto\Roth^{1/3}\).  Nevertheless, \(\deltah\sin\maxlat\) likely does indirectly affect the slantwise convective neutrality by influencing \(s_\mr{b}\).


Separately, F17 show that the latitude of the ITCZ, defined as the latitude of maximum zonal-mean precipitation, in their seasonal-cycle simulations scales as \(\Omega^{-0.63}\), very close to the \(\Omega^{-2/3}\) scaling predicted by \eqref{eq:eta-term3-approx2} (\cf/ their Fig.~6).   This ITCZ latitude is equatorward of the cell edge (c.f. their Fig.~5) and could in principle exhibit a unique scaling with \(\Roth\) to that of the cell edge.  Instead, evidently these hydrological and dynamical markers of the ascending branch position vary in proportion to one another and, in turn, the supercritical forcing extent.

\section{Summary}
\label{sec:summ}

We have presented a new theory for the latitude of the ascending edge of Earth's Hadley circulation during solsticial seasons and tested the theory's predictions against simulations in idealized GCMs.  The theory posits that the ascending edge latitude is determined by the meridional extent of supercritical forcing.  A supercritically forced latitude is one at which, supposing no large-scale overturning circulation existed, the resulting state of latitude-by-latitude RCE would generate time-mean distributions of angular momentum and/or absolute vorticity that are impossible.  It directly follows that a large-scale circulation must exist that spans at the very least all latitudes that are supercritically forced.  The resulting overturning circulation, however, can and typically does span poleward of this lower bound, leading to our empirical ansatz that the ascending edge latitude is proportional to the supercritical forcing extent.  Despite this empiricism, we argue that the resulting theory --- which is predictive and largely accurate with respect to the simulations we test it against --- offers advantages over other existing theories relevant to the problem.

We use a single-column model to simulate RCE at individual latitudes under Earth's present-day solsticial insolation, and by concatenating the simulations together we infer gradient-balanced zonal wind, angular momentum, and absolute vorticity distributions.  We then use a simple analytical expression (\ref{eq:lh88-forcing}), originally from LH88, to approximate the simulated RCE depth-averaged temperature field as quadratic in \(\sin\lat\) with its maximum located in the summer hemisphere.  The resulting expression for the absolute vorticity zero crossing, \ie/ the supercritical forcing extent, can be solved analytically in the Earth-relevant limit.  The solution states that the ascending edge latitude is proportional to \(\Roth^{1/3}\).  The solution is also unchanged if the RCE depth-averaged temperatures vary linearly in \(\sinlat\) (or any positive integer power in \(\sinlat\), for that matter) rather than quadratically.  This indicates that in the Earth-like regime the dominant influence on the supercritical forcing extent is the linear portion of the forcing in \(\sin\lat\), \ie/ the overall increase from the equator toward the summer mid-latitudes.

We examine the ascending edge latitude in simulations performed in two variants of an idealized, moist GCM and an idealized dry GCM, across each of which planetary rotation rate is varied.  Under solsticial conditions, in each model the cross-equatorial Hadley cell expands meridionally as the rotation rate decreases, and for diagnosed \(\Roth\) values up to order-unity, this expansion follows the \(\Roth^{1/3}\) scaling predicted by our approximate solution.  Simulations with very slow rotation rates and thus large \(\Roth\) values deviate from the scaling, but in a way that qualitatively resembles the more general solution (solved numerically).

We do not rest satisfied with a theory whose accuracy is qualitative, whose justification is semi-empirical, and whose strict interpretation is as a lower bound rather than a precise prediction.  We do consider it a useful step forward.

%
\acknowledgments
We are very grateful to Sean Faulk and Martin Singh for sharing the data from their simulations and for many valuable discussions.
S.A.H. acknowledges financial support from NSF award \#1624740 and from the Monsoon Mission, Earth System Science Organization, Ministry of Earth Sciences, Government of India.
J.L.M. acknowledges funding from the Climate and Large-scale Dynamics program of the NSF, award \#1912673.  We thank Martin Singh and two anonymous reviewers for helpful comments.


\appendix[A]
\appendixtitle{Choice of free parameters in the LH88 forcing approximation}
\label{app:free-params}
\setcounter{figure}{0}

For a wide range of \(\maxlat\) values spanning from the subtropics to the summer pole, reasonably accurate approximations to the numerical RCE simulations (at least with respect to the fields of relevance to supercritical forcing) can be found by tuning the value of \(\deltah\).  We perform a two-dimensional parameter sweep of (\ref{eq:lh88-forcing}), for \({1^\circ\leq\maxlat\leq90^\circ}\) in 0.1\degr{} increments and \({0.01\leq\deltah\leq0.3}\) in 0.01 increments.  For each profile, we compute \(\pdsl{\hat\theta\rce}{\lat}\) and compare it to the corresponding \(\pdsl{\hat T}{\lat}\) value from the numerical RCE simulations over the latitudes 45\degr{}S-45\degr{}N, selecting for each \(\maxlat\) the \(\deltah\) value that minimizes the root mean square error.

Fig.~A1 summarizes the results of these calculations, showing as a function of \(\maxlat\) the minimum root mean square error, the corresponding \(\deltah\) value, the corresponding value of the product \(\deltah\sin\maxlat\), and the corresponding supercritical extent.  The error in the analytical meridional temperature gradient field relative to the simulated one over 45\degr{}S-45\degr{}N is minimized for \(\maxlat=36^\circ\) with \(\deltah\approx0.145\approx1/7\).  Moving equatorward thereof, the best-fit \(\deltah\) increases, and the error metric increases considerably.  Moving poleward thereof, the best-fit \(\deltah\) decreases, and the error metric levels off at only slightly higher values.

\begin{figure}[t]
  \noindent
  \includegraphics[width=0.5\textwidth,angle=0]{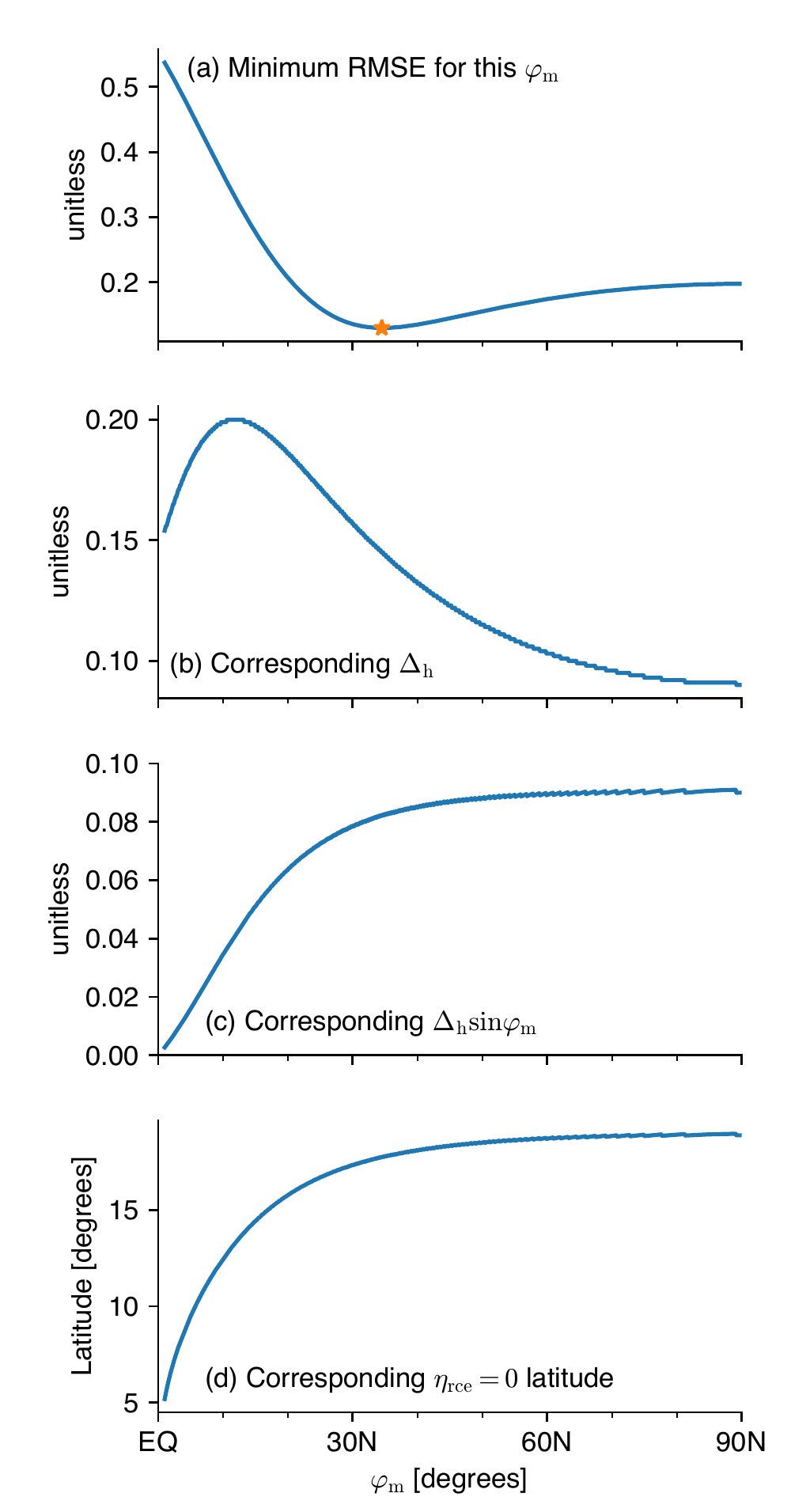}\\
  \appendcaption{A1}{Results from two-dimensional parameter sweep of (\ref{eq:lh88-forcing}), in \(\maxlat\) and \(\deltah\), with respect to the accuracy of the fit to the meridional temperature derivative field over 45\degr{}S-45\degr{}N from the numerical simulations of solsticial RCE.  Panel a shows the minimum root-mean-square error (RMSE) obtained as a function of \(\maxlat\).  Panel b shows the \(\deltah\) value corresponding to that minimum RMSE value.  Panel c shows the product \(\deltah\sin\maxlat\) using those values.  Panel d shows the latitude where \(\etarce=0\) using those values.}
  \label{fig:lh88-sweep}
\end{figure}

This decrease in the best-fit \(\deltah\) value as \(\maxlat\) is increased leads to the product \(\deltah\sin\maxlat\) remaining remarkably constant across the profiles with \(\maxlat\geq36^\circ\).  This is important, because \(\sin\maxlat\) only appears multiplied by \(\deltah\) in the analytical expressions shown below for the supercritical forcing extent (though \(\deltah\) separately appears on its own).  In other words, the LH88 approximations to the true RCE state, which might otherwise seem degenerate in \(\maxlat\) and \(\deltah\), effectively collapse into a single solution in \(\deltah\sin\maxlat\) space, at least with respect to the supercritical forcing extent.

\appendix[B]
\appendixtitle{Absolute vorticity zero crossing for \(\hatthetarce\) being an arbitrary polynomial in \(\sinlat-\sinmaxlat\)}
\label{app:polynomial}
\setcounter{figure}{0}

Let the RCE depth-averaged potential temperature field take the form
\begin{equation}
  \label{eq:theta-polynomial}
  \dfrac{\hat\theta}{\theta_0}=c_0-c(\sinlat-\sin\maxlat)^n,
\end{equation}
where \(n\) is a positive integer and \(c_0\) and \(c\) are constants.  For example, \eqref{eq:lh88-forcing} is the special case of (\ref{eq:theta-polynomial}) with \(n=2\), \(c_0=1+\deltah/3\), and \(c=\deltah\).  Using (\ref{eq:theta-polynomial}) with (\ref{eq:u-rce-lh88}), (\ref{eq:ang-mom-lh88}), and (\ref{eq:abs-vort-lh88}) yields the corresponding gradient-balanced zonal wind, absolute angular momentum, and absolute vorticity fields.  After introducing \({\tilde R\equiv c\Bu}\) (in analogy to \({\Roth=\deltah\Bu}\)), and for notational compactness \({\mu\equiv\sinlat}\) and \({\mumax\equiv\sinmaxlat}\), these are
\begin{equation}
  \label{eq:u-polynomial}
  u=\Omega a\coslat\left[\sqrt{1-n\tilde R\dfrac{(\mu-\mumax)^{n-1}}{\mu}}-1\right],
\end{equation}
\begin{equation}
  \label{eq:ang-mom-polynomial}
  M=\Omega a^2\cos^2\lat\sqrt{1-n\tilde R\dfrac{(\mu-\mumax)^{n-1}}{\mu}},
\end{equation}
and
\begin{align}
  \label{eq:abs-vort-polynomial}
  \eta&=2\Omega\sinlat\sqrt{1-n\tilde R\dfrac{(\mu-\mumax)^{n-1}}{\mu}}\times\\\nonumber
  &\left[1+\dfrac{n\tilde R}{4}\cos^2\lat\dfrac{(\mu-\mumax)^{n-2}}{\mu^2}\dfrac{(n-2)\mu+\mumax}{\mu-nR(\mu-\mumax)^{n-1}}\right].
\end{align}
Setting the last, square-bracketed term of (\ref{eq:abs-vort-polynomial}) equal to zero yields, after some manipulation,
\begin{align}
  \label{eq:abs-vort-zero-poly}
  &\mu^3-n\tilde R\mu^2(\mu-\mumax)^{n-1}\\\nonumber
  &+\dfrac{n\tilde R}{4}\cos^2\lat\left[(n-2)\mu-\mumax\right](\mu-\mumax)^{n-2}=0.
\end{align}

Now consider the small-\(\lat\), small-\(\tilde R\) limit.  Without loss of generality, we can set \(\mumax=1\), because as described in Section~\ref{sec:rce-sims} for the \({n=2}\) case, an accurate fit to the actual solsticial insolation profile can be found for any extratropical \(\maxlat\) value by adjusting the value of \(c\).  We then have
\begin{equation}
  \label{eq:abs-vort-zero-poly-small-angle}
  \lat^3-n\tilde R\lat^2(\lat-1)^{n-1}+\dfrac{n\tilde R}{4}[(n-2)\lat-1](\lat-1)^{n-2}=0.
\end{equation}
The left hand side comprises the sum of three terms.  In the \(\tilde R\ll\lat\ll1\) limit considered in the main text for the \(n=2\) case, to lowest order the three terms are of magnitude \(\lat^3\), \(\tilde R\lat^{n+1}\), and \(\tilde R\), respectively.  Since \(\tilde R\ll\lat\), for \(n\geq1\) we have \(\tilde R\lat^2\ll\lat^3\), and therefore the leading order balance is between the first and third terms:
\begin{equation}
  \label{eq:zero-poly-approx}
  \lat\approx\left(\dfrac{n\tilde R}{4}\right)^{1/3}.
\end{equation}


\appendix[C]
\appendixtitle{Formulation of idealized GCMs used}
\label{app:gcms-desc}
\setcounter{figure}{0}

The simulations of F17 were performed in the \citet{frierson_gray-radiation_2006} idealized aquaplanet GCM.  This model's spectral dynamical core solves the primitive equations on the sphere with no topography and a water-covered surface.  The sigma vertical coordinate is defined according to the local surface pressure, \(\sigma=p/p_\mr{s}\).  Simplified gray radiative transfer is used with a prescribed, time-invariant, meridionally uniform longwave optical depth field, no shortwave absorption in the atmosphere, and a prescribed, uniform surface albedo.  Surface turbulent fluxes of latent heat and sensible heat are calculated via standard bulk aerodynamic formulae.  The surface approximates the thermodynamic effects of the ocean's upper, well-mixed layer.  Its temperature tendency is determined by the net downward radiative plus turbulent flux into the surface along with the prescribed heat capacity, which corresponds to a water depth of 10~m.  There is no prescribed ocean heat flux divergence (\ie/ ``Q-flux'').

Moist convection is parameterized using the convective adjustment scheme of \citet{frierson_dynamics_2007}, based on so-called Betts-Miller schemes \citep{betts_new_1986-1,betts_new_1986}, that relaxes the humidity and temperature profiles of convectively unstable columns toward a moist adiabat with a prescribed 70\% relative humidity over a fixed 2-hr timescale.  Neither water vapor nor cloud radiative feedbacks operate, the former because the prescribed longwave optical depth field does not depend on water vapor.  The latter is because there are no clouds --- liquid water generated either through the convective parameterization or by grid-scale saturation is immediately precipitated out to the surface.

We refer readers to F17 and \citet{frierson_gray-radiation_2006} for further details on the model formulation.  We refer readers to S19, \citet{frierson_gray-radiation_2006}, and \citet{ogorman_hydrological_2008} for further details on the model formulation.

We perform additional simulations in the dry idealized GCM of \citet{schneider_tropopause_2004}.  This model uses the same spectral dynamical core as the moist simulations just described, with horizontal resolution T85 and 20 unevenly spaced sigma levels.  The vertical dependence of its Newtonian relaxation temperature field approximates the radiative equilibrium temperature profile of a semi-gray atmosphere in the troposphere, and it more crudely represents the stratosphere as an isothermal layer of 200 K extending to the model top.  The Newtonian relaxation timescale is 50~days in the free atmosphere, 7~days at the surface, and varies linearly in \(\sigma\) within the planetary boundary layer with prescribed top at \(\sigma=0.85\).

Within the troposphere, the equilibrium temperature profile is statically unstable over much of the troposphere, and at each timestep any statically unstable column triggers a convective adjustment procedure.  The convective adjustment relaxes statically unstable columns over a uniform 4-day timescale toward a prescribed lapse rate of \(\Gamma=\gamma\Gamma_\mr{d}\), where \(\Gamma\) is the lapse rate, \(\Gamma_\mr{d}=g/c_p\) is the dry adiabatic lapse rate, and \(\gamma=0.7\).  The \(\gamma\) term acts to mimic the stabilizing effects of latent heat release by moist convection while retaining the simplicity of an otherwise dry fluid.  The two dissipative processes are a conventional \(\nabla^8\) hyperdiffusion and a quadratic drag on the zonal and meridional winds within the boundary layer.  Additional details of the model formulation are described by \citet{schneider_tropopause_2004}, and note that various additional modifications made by \cite{hill_axisymmetric_2019} --- in particular making the model axisymmetric --- are not employed in the present study.

For all simulations, we diagnose the ascending edge latitude as follows.  Denoting the meridional mass overturning streamfunction \(\Psi(\lat,\sigma)\), its maximum value above the boundary layer (\ie/ at the Hadley cell center) \(\Psi_\mr{max}\), and the sigma level and latitude of \(\Psi_\mr{max}\) as \(\sigma_\mr{max}\) and \(\lat_\mr{max}\) respectively, \(\ascentlat\) is the latitude in the summer hemisphere satisfying
\begin{equation}
  \label{eq:edge-def}
  \dfrac{\Psi(\ascentlat,\sigma_\mr{max})}{\cos\ascentlat}=\alpha\dfrac{\Psi_\mr{max}}{\cos\lat_\mr{max}},
\end{equation}
where \(\alpha=0.1\).  Apart from the cosine factors, this is equivalent to the standard edge definition based on where the streamfunction drops below the specified fraction \(\alpha\) (set here, as typical to 0.1) of its maximum value at the level of that maximum \citep[\eg/][]{walker_eddy_2006}; using a small but nonzero fractional threshold is needed for cases in which a non-global Hadley cell emerges, but a Ferrel cell does not, leading to the streamfunction retaining its sign all the way to the pole.  The cosine terms act as weights accounting for the decreasing circumference of latitude circles moving poleward.  It yields cell edges farther poleward than the conventional definition, the more so the larger the cell, but results are qualitatively insensitive to whether this weighting is applied (not shown).








%
%
%
\bibliographystyle{ametsoc2014}
\bibliography{./references}

%

%

\end{document}